\documentclass[letterpaper, twocolumn, superscriptaddress, showpacs, preprintnumbers, amsmath, amssymb,longbibliography, nofootinbib]{revtex4-1}
\usepackage[pdftex]{graphicx}
\usepackage{caption}
\usepackage{subfigure}
\usepackage{dcolumn}
\usepackage{bm}
\usepackage{color}
\usepackage{hyperref}
\usepackage{adjustbox}
\usepackage{color}
\usepackage{float}
\usepackage{framed}
\usepackage{esvect}
\usepackage{amsfonts,amssymb,amsmath,hyperref,hypcap,enumerate}
\usepackage{wasysym}
\usepackage{slashed}
\usepackage{tikz}
\usepackage{amsmath,amssymb,bm,graphicx,dsfont}
\usepackage[latin9]{inputenc}
\usepackage{amsmath}
\usepackage{slashed}
\usepackage{dsfont}
\usepackage{amstext}
\usepackage{amssymb}
\usepackage{amsbsy}
\usepackage{amsthm}
\usepackage{epsfig}
\usepackage{tikz}
\usepackage{bbm}
\usepackage{hyperref}
\usepackage{color}
\usepackage{multirow}
\usepackage{ulem}
\usepackage{soul}
\usepackage[T1]{fontenc}
\newcommand{\be}{\begin{equation}}
\newcommand{\ba}{\begin{align}}
\newcommand{\ee}{\end{equation}}
\newcommand{\bea}{\begin{eqnarray}}
\newcommand{\eea}{\end{eqnarray}}
\newcommand{\beq}{\begin{equation}}
\newcommand{\eeq}{\end{equation}}
\newcommand{\beqn}{\begin{eqnarray}}
\newcommand{\eeqn}{\end{eqnarray}}

\renewcommand{\vec}[1]{{\bf #1}}
\renewcommand{\hat}[1]{{\widehat #1}}

\begin{document}
\title{
Doping a moir\'e Mott Insulator: A t-J model study of twisted cuprates }
\author{Xue-Yang Song}
\author{Ya-Hui Zhang}
\author{Ashvin Vishwanath}
\affiliation{Department of Physics,  Harvard University,Cambridge, MA 02138, USA}
\date{\today}

\begin{abstract}
We theoretically investigate twisted structures where each layer is composed of a strongly correlated material. In particular, we study a twisted t-J model of cuprate multilayers  within the slave-boson mean field theory. This treatment encompasses the Mott physics at small doping and self consistently generates d-wave pairing. Furthermore, including the  correct inter-layer tunneling form factor consistent with the symmetry of the Cu $d_{x^2-y^2}$ orbital proves to be crucial for the phase diagram. We find spontaneous time reversal (T) breaking around twist angle of $45^\circ$, although only in a narrow window of twist angles. Moreover, the gap obtained is small and the Chern number vanishes, implying a non-topological superconductor. 
At smaller twist angles, driving an interlayer current however can lead to a gapped topological phase. The energy-phase relation of the  interlayer Josephson junction displays notable double-Cooper-pair tunneling which dominates around $45^o$. The twist angle dependence of the Josephson critical current and the Shapiro steps are consistent with  recent experiments. Utilizing the moir\'e structure as a  probe of correlation physics, in particular of the pair density wave state, is discussed.
\end{abstract}
\maketitle

\emph{Introduction-} Twisted heterostructures of graphene and  transition metal dichalcogenides (TMDs) have attracted significant attention for displaying a series of interaction-driven phenomena such as superconductivity  \cite{cao_pablo,cao182},  integer and fractional Chern insulator phases \cite{ggordon,young19,efetov,xie, Mak21} and as simulators of the Hubbard model   \cite{MacDonald,tang_2019,Regan2019optical,wang2019magic}. Although the single layers tend to be weakly correlated, the twist structure reconstructs the electronic bands  creating a versatile platform to study strongly interacting physics.  A natural next step towards even richer phenomena are  twisted bilayers where each layer  itself is  strongly correlated.  A promising candidate is twisted cuprate bilayers.

The recent fabrication of high quality monolayer of $\textrm{Bi}_2\textrm{Sr}_2\textrm{CaCu}_2\textrm{O}_8$ (Bi2212)\cite{monolayer} finds similar critical temperature in single-layer and bulk cuprates, opening up the experimental study of cuprates as essentially $2D$ system. This also enables the experimental study of twisted cuprate bilayer, which in turn has spurred pioneering theoretical proposals for two different realizations of topological superconductor in twisted cuprate bilayer\cite{franz,pixley20}. Especially, Ref.~\onlinecite{franz} proposes a topological superconductor with time reversal symmetry spontaneously broken when the twist angle $\theta$ is around $45^\circ$ degree while Ref.~\onlinecite{pixley20} seeks to stabilize a similar state by tuning near a magic angle with weak quasiparticle dispersion and establishing a superflow along the `c-axis'. On experimental side, twisted cuprates have recently been fabricated and characterized \cite{QKXue,zhao21}. Ref \cite{zhao21} reported a measurable  second harmonic Josephson coupling at $\theta \approx 45^\circ$ based on the Shapiro step. However, so far there is no experimental signature of gap opening,spontaneous time-reversal symmetry breaking or  topological superconductivity. 

 Here we study the twisted cuprate bilayer problem  theoretically, augmenting earlier studies in two significant ways (I) we introduce a microscopic twisted t-J model and solve it within the self-consistent slave-boson mean-field approximation, which accounts for strong correlations. The $d$ wave pairing ansatz is obtained self consistently. In contrast, in a BCS approach one incorporates  attractive interaction in the d wave channel by hand \cite{franz} . (II) We incorporate the Mott physics in the slave boson mean field theory. Such a method is known to be successful in capturing the d-wave superconductor in single layer\cite{rmp_lee}. In twisted bilayer, the effective inter-layer tunneling will be suppressed by a factor $x$ in this approach, where $0< x<1$ is the hole doping level. Hence the method can smoothly crossover to the $x=0$ limit which represents the  undoped Mott insulator where the inter-layer tunneling should be fully suppressed. This $x$ factor makes the window of twist angle with time reversal breaking and the gap significantly smaller than that within a BCS treatment \cite{franz}. (III) We note that the inter-layer tunneling between the two cuprate layers should include a form factor $(\cos k_{x,t}-\cos k_{y,t})(\cos k_{x,b}-\cos k_{y,b})$\cite{cupratetunnel}, where the momentum $\vec{k}_t$ and $\vec{k}_b$ are defined in the top and bottom layer separately, with the coordinate system fixed to the Cu-O bond of each layer. 
 
 Such a form is especially important to consider since it should vanish at $\theta=45^\circ$ by symmetry, or more specifically due to the symmetry properties of the Cu $d_{x^2-y^2}$ orbital\footnote{Strictly speaking, the oxygen $p$ orbital is also involved in hole doped cuprate, forming the Zhang-Rice singlet. However, the active Zhang-Rice singlet has the same symmetry as the Cu $d_{x^2-y^2}$ orbital and  does not change our analysis.}. We note that the form factor suppresses the tunneling around the nodal region and therefore the Dirac nodes remain almost gapless even if  time reversal symmetry is  broken due to the pairing phase difference between the two layers. Incorporating such momentum dependent tunneling terms will lead to key differences from  earlier studies that employ a momentum independent tunneling term \cite{franz}.

 Our work has important implications for the ongoing experimental studies of twisted cuprate bilayer\cite{zhao21}. We extract the inter-layer Josephson coupling $E[\phi]=-a \cos \phi+b \cos 2\phi$ for different twist angle $\theta$, where $\phi$ is the phase difference between the two layers. Then we can extract the critical Josephson current and  demonstrate similar twist angle dependence as observed in experiment Ref. \cite{zhao21}. In particular we find $|a|<b$ at $\theta=45.2^\circ$ and calculate the Shapiro steps at this angle. The results are consistent with the Shapiro steps measured in Ref.~\onlinecite{zhao21}. Near this angle,  we observe spontaneous time reversal breaking due to $\phi \neq 0$ as reported in earlier studies \cite{franz}. However, in contrast to the earlier studies, due to the additional ingredients in our treatment, we find no significant gap nor topological chiral modes.  Instead, we calculate the polar Kerr effect\cite{kapitulnik,cankerr} which can be used as a diagnostic of spontaneous time reversal  breaking and propose future optical experiments to verify this effect. 
 

However, following Ref. \cite{pixley20}, we do find that topological superconductivity (SC) can be induced by adding an interlayer supercurrent by hand to give $\phi \neq 0$. This is  despite the fact that the magic angle is suppressed due to the $x$ factor in the inter-layer tunneling in slave boson theory. This again highlights the importance of Mott physics in modeling twisted correlated bilayers. 


Finally, we also point out an interesting application of the moire' heterostructure as a {\it probe} of cuprates, in particular the pair density wave physics, that has been discussed extensively \cite{agterberg2020physics,himeda2002stripe,berg2007dynamical,berg2009striped,fradkin2015colloquium,lee2014amperean,edkins2019magnetic}. 

\emph{Model for twisted double-bilayer Bi2212-}
We take the $t-J$ model to describe a single cuprate layer on square lattice:
\begin{equation}
\label{tjh}
\mathcal H_{tJ}=-t\sum_{\langle ij\rangle} Pc_{i,s}^\dagger c_{j,s}P+J( \mathbf S_i\cdot \mathbf S_j-\frac{1}{4}n_in_j)+\mu\sum_i (n_i-1),
\end{equation}
where $s=\uparrow,\, \downarrow; \, n_i=\sum_s c_{i,s}^\dagger c_{i,s}$, $\mathbf S=\frac{1}{2} c_s^\dagger \mathbf \sigma_{ss'} c_{s'}$ and $P$ projects out states with  double/zero occupancy on any site for hole/electron doping, respectively. We take $t=2J=0.2eV$ in the calculation. Note that a transformation $c_{is}\rightarrow c_{is}^\dagger$ relates electron to hole doping cases, with the sign of $t,\mu$ changed. Here we focus on hole doping and leave electron doping for future studies.

\begin{figure}
 \captionsetup{justification=raggedright}
    \centering
        \adjustbox{trim={.04\width} {.2\height} {.25\width} {.1\height},clip}
    {\includegraphics[width=.65\textwidth]{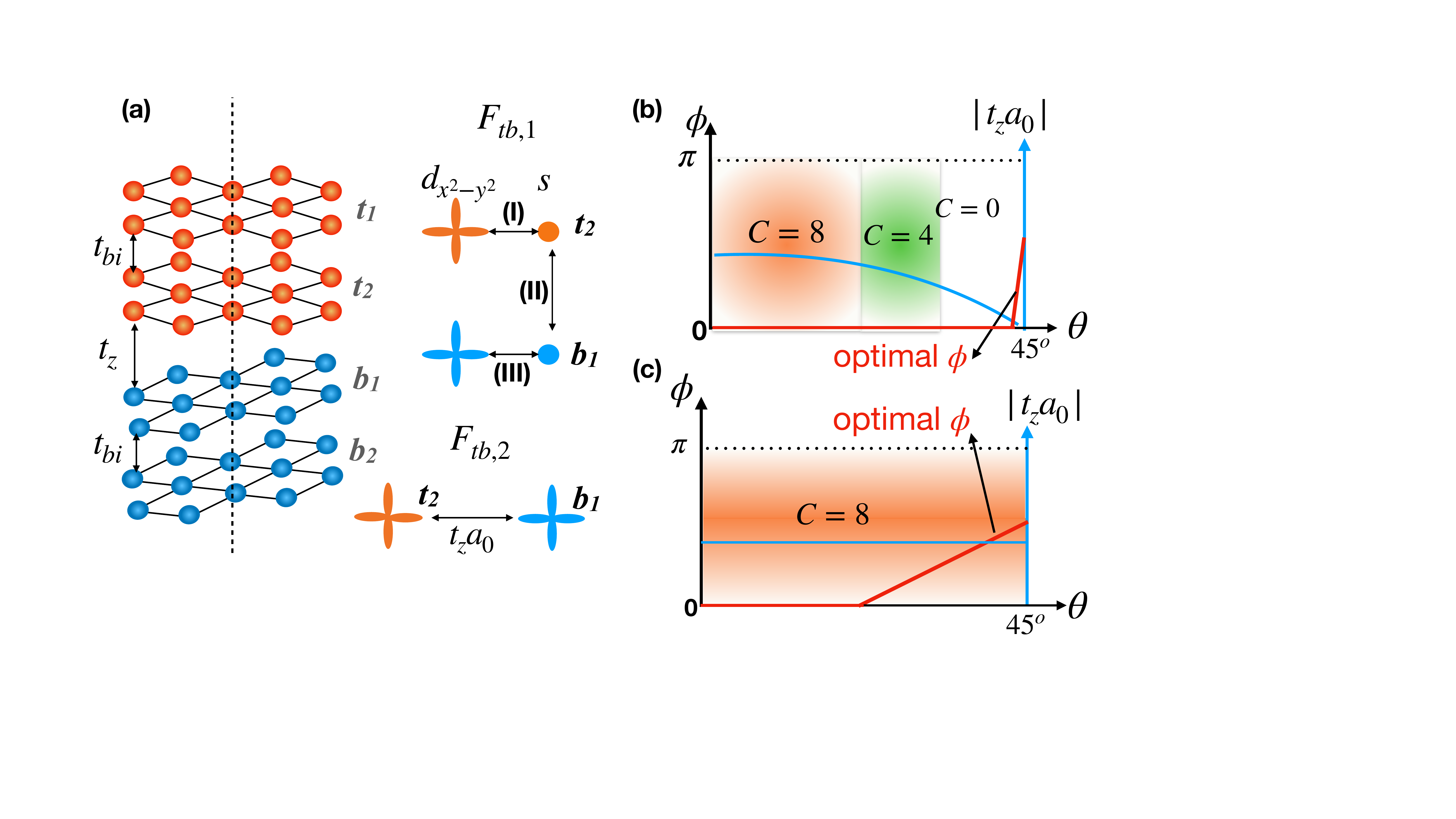}}
    \caption{{\bf (a)} A schematic plot of the twisted $4$ CuO layers. The sites with the same color within the top, bottom bilayers are perfectly aligned. There is an hopping across the bilayer within unit cell with strength $t_{bi}$ and interlayer hopping with strength $t_z$. Tunneling across the twist consists of $2$ processes: $F_{tb,1}$ is a $3$ step tunneling process mediated by $s$ orbital of $Cu$ and $F_{tb,2}$ describes a direct tunneling between $d$ orbitals of $Cu$. {\bf (b,c)} The schematic phase diagram of twisted cuprates as one varies twist angle $\theta$ and pairing phase difference $\phi$ between top and bottom layers. The two figures present cases with {\bf (b)} mixed tunneling i.e. both $\cos$ form and uniform tunneling which vanishes at $\theta=45^o$.
    The Chern numbers for small twist angles are shown.
    Other Chern numbers e.g. $C=2,6$ are realized with different tunneling strength; spontaneous breaking of time reversal is shown by the red line, when it separates from the  x-axis, revealing a nonzero optimal $\phi$ {\bf (c)} only uniform tunneling, results are comparable to Ref. \cite{franz}. Actual data of $\phi$ vs $\theta$ are plotted in fig \ref{fig:phiuniform}. }
    \label{fig:twist}
\end{figure}

Motivated by experiment \cite{QKXue,zhao21} we consider twisted {\em double} bi-layers, and add an index $p=t1,t2,b1,b2$ denoting operators in the top first, top second, bottom first and bottom second layers, as shown in fig \ref{fig:twist}.The twist by angle $\theta$ occurs between layers $t2$ and $b1$. Each layer is hence defined as top (bottom) bilayers.  The full Hamiltonian should include interlayer tunneling with strength $t_z$ as,
\begin{eqnarray}
\label{tbsc}
\mathcal H_{tbsc}=\sum_{i=1,2}\mathcal H_{tJ,ti}+\mathcal H_{tJ,bi}
-t_z\sum_{ij}F_{tb}(\vec r_{ij}) c^\dagger_{i,t2} c_{j,b1}\nonumber\\
-t_{bi}\sum_{n,p=t,b}c_{n,p1}^\dagger c_{n,p2} +h.c.,
\end{eqnarray}
where $t_z$ term describes interlayer tunneling which we will come back to later. The second line describes tunnel between layers inside top (bottom) bilayers (i.e. within unit cell) and $n$ labels sites inside a Moire unit cell in a monolayer. The function $F_{tb}$ is the form factor for inter-layer tunneling resulting from the orbital characteristics of $Cu$ ions. 

Interlayer tunneling between $d$ orbitals primarily consists of two processes, plotted in Fig. \ref{fig:twist}(a): The first one is mediated by $s$ orbitals and the second one is a direct tunneling controlled by the dimensionless parameter $a_0$\cite{Bansil}\footnote{$a_0=1.6$ in untwisted cuprates (appendix \ref{bandstr})\cite{Bansil}.}.The two processes give two factors, to wit
\begin{eqnarray}
    \label{tbform}
    F_{tb,1}(\vec r_{ij})&=&\frac{1}{2}\sum_{\hat h=x,y,\hat h'=x',y'} \xi_{\hat h}\xi_{\hat h'} g_s(r_{i\pm\hat h,j\pm\hat h'}),\nonumber\\    
    \xi_{\hat x}=1,\,\xi_{\hat y}=-1,\,&&g_s(i_{t2},j_{b1})=e^{-(l_{ij}-l_d)/\rho_s},\nonumber\\
F_{tb,2}(\vec r_{ij})&=&a_0 g_s(i_{t2},j_{b1}) ,
\end{eqnarray}
where the first process has a form in momentum space $F_{tb,1}(\vec k)\propto (\cos k_{x,t}-\cos k_{y,t})(\cos k_{x,b}-\cos k_{y,b})$ (subscript $t,b$ denotes momentum in the relative canonical coordinates of top, bottom layers, respectively), descending from the form $(\cos k_x-\cos k_y)^2$ in untwisted cuprates\cite{cupratetunnel}.  Here $l_{ij} = |i_{t2}-j_{b1}|$ and $l_d\approx 2a_{Cu}$ ($a_{Cu}\approx 0.3nm$ the lattice constant of $CuO$ plane) is the distance between layer, and $\rho\approx 0.5a_{Cu}$ is a tunneling parameter. 

From a symmetry $R_{xy}$ that reflects along a diagonal direction of one of the cuprate layer, one deduces 
\begin{equation}
    a_0(\theta)=-a_0(90^o-\theta)\nonumber\\,
    a_0(45^o)=0.
\end{equation}
Details on tunneling forms, symmetry constraints and parameter choices from band structures of Bi2212 are listed in appendix \ref{symmetry}, \ref{bandstr}, respectively.

Vanishing of $a_0$ close to $45^o$ results in qualitative change of phase diagram for twisted cuprates plotted in fig \ref{fig:twist}(b,c), e.g. chiral non-topological superconductivity(SC) and smaller gap etc, compared to finite $a_0$ (details in app \ref{comparebcs}).


\emph{Slave boson mean field theory and self-consistent solutions-}
To decouple the Hamiltonian we adopt a parton decomposition for electron operators,
\begin{equation}
c_{i,s}=b_i^\dagger f_{i,s}
\end{equation}
with fermionic spinons $f_s$ and bosonic holons $b$. There is a gauge constraint $\sum_s f^\dagger_{i,s}f_{i,s}+b_i^\dagger b_i=1$ for states excluding double occupancy, and the electron number reads
\begin{equation}
n_i=1-b_i^\dagger b_i= 1-x_i.
\end{equation}

 Upon doping with a fraction $x$ of holes, the condition becomes $\sum_i b_i^\dagger b_i=Nx$ with $N$ being the number of sites. The holons will condense for nonzero $x>0$ to $\langle b_i\rangle=\sqrt{x_i}$. Substituting the condensed holon operator into the kinetic terms with hopping $t$ and tunneling $g$ in the twisted bilayer $t-J$ Hamiltonian eq\eqref{tbsc}, and further decouple the  $\mathbf S_i\cdot \mathbf S_j$ by a mean-field treatment\cite{slave_tj}, one gets (assuming spin rotation invariance)
 \begin{eqnarray}
 \label{mfh}
 \mathcal H_{MF}&=&\sum_{p=t(b)}\{-\sum_{\langle ij\rangle, s} [(t\sqrt{x_{ip}x_{jp}}+\frac{3J}{8} \chi_{ij,p}^*)f_{isp}^\dagger f_{isp}+h.c.]\nonumber\\
 -&&\frac{3J}{8}\sum_{\langle ij\rangle ss'} [\Delta_{ij,p}^* f_{is,p}f_{js',p}\epsilon_{ss'}+h.c.]+\mu_p\sum_i n_{ip}\}\nonumber\\
 -&&t_z\sum_{ij,x}F_{tb}(\vec r_{ij}) e^{-(l_{ij}-l_d)/\rho}\sqrt{x_{i,t2}x_{j,b1}} f^\dagger_{is,t2} f_{js,b1}\nonumber\\
-&&t_{bi}\sum_{k,p=t,b}F_p(k)\sqrt{x_{n,p1}x_{n,p2}}f_{kn,p1}^\dagger f_{kn,p2} +h.c.,
 \end{eqnarray}
 where we mainly consider the mean field order parameters for hopping $\chi_{ij}=\sum_s \langle f_{is}^\dagger f_{is}\rangle$, pairing $\Delta_{ij}=\sum_{ss'} \epsilon_{ss'}\langle f_{is}f_{js'}\rangle$, and ignore local magnetic moments. 
 The interlayer tunneling $F_{tb}=F_{tb,1}+F_{tb,2}$ consists of $2$ processes shown in eqs \eqref{tbform}. We use $t_za_0$ to mark the uniform tunneling strength hereafter and from \cite{Bansil} in untwisted Bi2212 $t_za_0\sim 0.16t$.
 
The parton construction naturally incorporates the Mott insulating state at vanishing doping $x$, since the system will enter a staggered flux spin liquid for $x\rightarrow 0$. Note that the eigenstate solutions are symmetric with respect to positive, negative energies due to the chiral symmetry $\mathcal C$ (appendix \ref{symmetry}). 
 
 We use $t/J=2$. For a mean-field treatment, we start with  a $d-$wave pairing ansatz, i.e. $\Delta _{i,i+\hat x}=-\Delta_{i,i+\hat y}=\delta_i (e^{i\phi})$. We emphasize this ansatz is not biased, i.e. it will be obtained even from random pairing ansatz.
In this way we automatically reach a mean-field minimum of the free energy.

We find that the pairing orders converge to a $d-$ wave form with a possible phase difference $\phi$ between $t,b$ layers,
\begin{eqnarray}
\Delta_{i,i+\hat x,t}=-\Delta_{i,i+\hat y,t}=\delta\nonumber\\
\Delta_{i,i+\hat x,b}=-\Delta_{i,i+\hat y,b}=\delta e^{i\phi},
\end{eqnarray}
with the phase difference $\phi\in (0,\pi)$.  We can also explicitly choose the ansatz with an arbitrary $\phi$ and calculate the energy $E(\phi)$. Details on numerics are listed in appendix \ref{app:num}.



\begin{figure}
 \captionsetup{justification=raggedright}
    \centering
                \adjustbox{trim={.08\width} {.02\height} {.1\width} {.05\height},clip}
   { \includegraphics[width=0.7\textwidth]{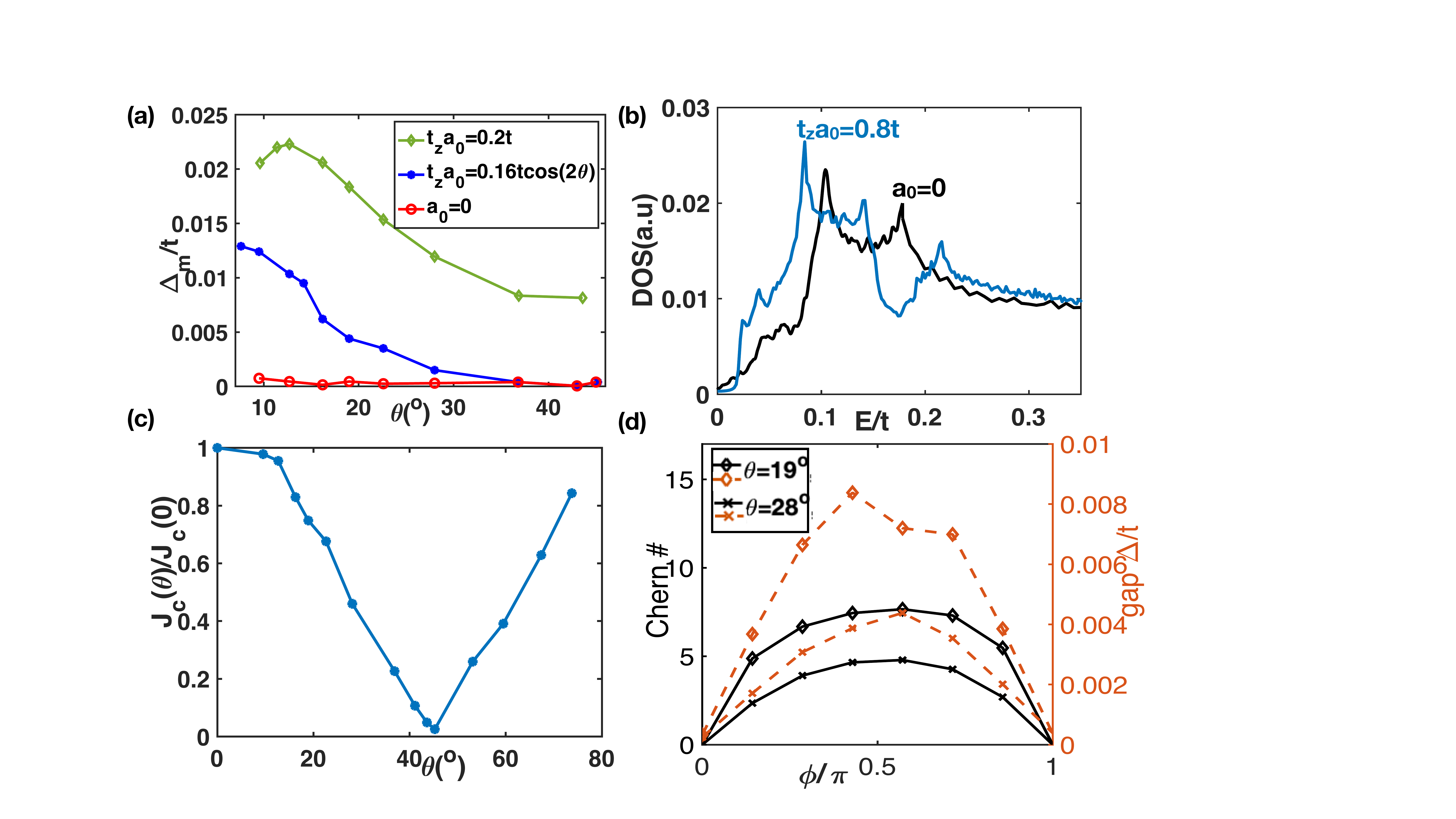}}
    \caption{ Data at $x=0.2,T=0$.{\bf (a)} The maximal gap $\Delta_m$, obtained by varying the inter-layer phase $\phi$ at fixed twist angle $\theta$, for different tunneling cases: Uniform tunneling only (green). Mixed tunneling, $\cos$ form factor mixed with uniform tunneling with realistic parameters (blue). Only $\cos$ form factor tunneling ($a_0=0$) (red), respectively. For the latter two $t_z=0.1t$.
    {\bf (b)} The density of states at $\theta=43.6^o$ for (only) uniform tunneling (blue) with a gap $\sim 4\% t$ and (only) $\cos$ form tunneling (black) with a vanishing gap $<0.001t$. The blue curve assumes a large uniform tunneling to make the gap easily discernible.
    {\bf (c)} Critical Josephson current density normalized by the untwisted value with $t_z=0.1t,\,a_0=0$.  The smallest $J_c$ is obtained at $\theta=45.2^o$ and has a value of $2000 kA/cm^2$, while $J_c(0^o)=8\times 10^4kA/cm^2$ with $J_c(45^o)/J_c(0^o)=0.025$.
    {\bf (d)} The Chern number (black) and gap (orange) upon varying $\phi$ for two relatively small twist angles with mixed tunneling and  $(t_z=0.1t,\,t_za_0=0.16t)$. 
    }\label{fig:phase_tunnel}
    \end{figure}
    
\begin{figure}
 \captionsetup{justification=raggedright}
    \centering
                \adjustbox{trim={.07\width} {.0\height} {.15\width} {.03\height},clip}
   { \includegraphics[width=0.7\textwidth]{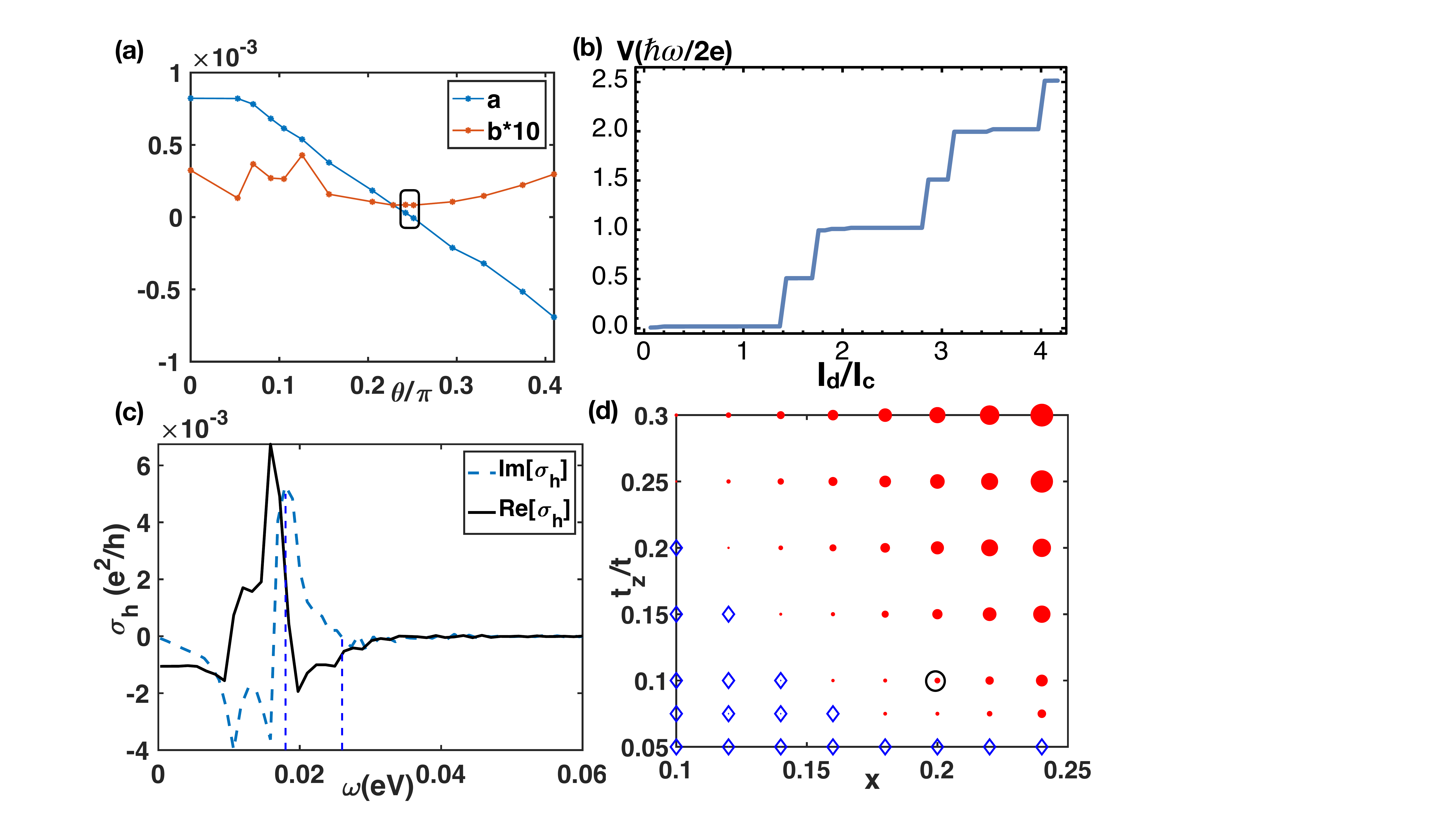}}
    \caption{Data for $t_z=0.1t, x=0.2,a_0=0$.{\bf (a)} The coefficients for fitting energy phase relation $E=-a\cos(\phi)+b\cos(2\phi)$ ($E$ is obtained per unit cell of $CuO$ planes), with vanishing uniform tunneling $a_0=0$. $a,b$ is plotted in units of $t/a_{Cu}^2$. 
    At $\theta=43.6,45.2^o$(boxed) we have an energy minima at nontrivial $\phi\approx 0.16\pi,0.64\pi $, respectively. 
    {\bf (b)} The Shapiro step measured in resistively shunted Josephson junction where a DC current $I_d$ and an AC current $I_{rf}$ with frequency $\omega$.
    The time-averaged voltage displays steps as $I_d$ is increased, in particular the half step at $V=0.5,1.5,2.5$ (in units of $\hbar \omega/2e$) indicates a second harmonic term in energy-phase relation. Parameters used are results for $\theta=45.2^o, b=-1.3a$, with $I_{rf}=1.25I_c,\omega=0.5I_c/e$ ($I_c$ is the critical Josephson current).
    {\bf (c)} Real and Imaginary parts of the Hall conductivity, $\sigma_h(\omega)$ at one particular parameter set $\phi=3\pi/16,a_0=0$ (circled in {\bf (d)}). The Kerr angle is estimated \cite{cankerr} to be of order $1\mu {\rm rad}$ for small frequency $\omega<0.03 eV$. The edge above which $\sigma_h$ tends to $0$ at $\omega/t\approx 0.2$ corresponds to the maximal gap around Fermi surface (fig \ref{fig:band11}) in untwisted cuprates. 
    {\bf (d)}The real part of anomalous Hall conductivity $Re[\sigma_h(\omega=0)]$ (proportional to the red disk area) at $\theta=43.6^o$ as one varies $x$ and $t_z$, signaling T breaking. The blue diamonds denote absence of T breaking, i.e. $Re[\sigma_h(0)]=0$.}
    \label{fig:tr_45}
    \end{figure}

\emph{Critical Josephson current-}
From the energy-phase $E$ vs $\phi$ relation (shown in figure \ref{fig:ene4f})\footnote{The energy variation upon varying $\phi$ comes mainly from interlayer tunneling terms and we calculate energy only from interlayer tunneling terms hereafter.}, one could extract critical Josephson current by $I=\frac{2e}{\hbar}max_{\phi}(\frac{dE}{d\phi})$ as shown in fig \ref{fig:phase_tunnel}(c). Fig \ref{fig:tr_45}(a) shows the fitting parameters $a,b$ as $E=-a\cos \phi+b\cos 2\phi+const$.

Experiments \cite{QKXue} on twisted Bi2212 yields a current density $J_c\approx 100A/cm^2$ at $\theta=45^o$, doping $x=0.1$. Another Josephson current measurement \cite{zhao21} shows strong angular dependence of $J_c$, with which Fig. \ref{fig:phase_tunnel}(c) qualitatively agrees, though \cite{zhao21} reports a smaller $J_c(45^o)\approx 40-120A/cm^2$.
 Numerically we find $J_c\approx 50kA/cm^2$ at $t_z=0.05t,x=0.1$, $500$ times larger than experiment. This reduction in critical current is puzzling, and we  conjecture that the vortex dynamics between top, bottom layers, rather than $t-J$ physics, determines $J_c$. We also find a significant $c$-axis magnetization, 
 as plotted in fig \ref{fig:mkz}.

\emph{$\theta\sim 45^o$: Chiral non-topological SC and anomalous Hall effects-}
We found, that the system energetically favors a nonzero $\phi$ and spontaneously breaks time-reversal symmetry close to $45^o$. Fig \ref{fig:tr_45}(a) shows the fitting parameters $a,b$ as $E=-a\cos \phi+b\cos 2\phi+const$. At $\theta=43.6^o,45.2^o$, we have $2|b|>|a|$ and an optimal $\phi\approx 0.16\pi,0.64\pi$, respectively. While for other angles, optimal $\phi=0,\pi$ for $\theta<41.2^o,>48.8^o$, respectively.

The spontaneous generation of time reversal breaking from $\phi \neq 0$ has already been pointed out previously\cite{franz}. However, the property of the chiral (T breaking) superconductor in our theory is qualitatively different. First, due to the $x$ factor appearing in the inter-layer tunneling in the slave boson theory, the window of twist angle with $\phi\neq 0$ is significantly smaller in our calculation. More importantly, the chiral superconductor shows a very small gap in our theory, because of the form factor in the inter-layer tunneling.  The form factor suppresses the inter-layer tunneling at nodal region and thus the gap opening is suppressed even if there is a large interlayer phase $\phi$. A significant uniform tunneling component, added by hand, will open a larger gap as plotted in fig \ref{fig:phase_tunnel}(a). The density of states(DoS) also shows qualitative difference at small energy between cases with or without uniform tunneling in fig \ref{fig:phase_tunnel}(b): DoS is V shaped(black) for vanishing gap with only $\cos$ form tunneling; DoS has a vanishing flat segment(blue) at low energy corresponding to the gap opened by uniform tunneling. We use a $4$ band continuum model to derive the gap in appendix \ref{continuum} to explain why $\cos$ form tunneling opens a much smaller gap than uniform tunneling at leading order. 

Besides, we find that the Chern number is zero, both for twisted double-bilayer and twisted bilayer Bi2212 (i.e., retaining only $t_2,b_1$ and interlayer tunnel, phase diagram in fig \ref{fig:phase2}). Therefore a topological superconductor is absent in our treatment, in contrast to Ref.~\onlinecite{franz}. If we ignore the form factor and just take a  homogeneous tunneling in our t-J calculation, we recovered results in Ref.\cite{franz} qualitatively, e.g. the gap, Chern number etc. (Details in appendix \ref{comparebcs}.) But unfortunately such an s-wave inter-layer tunneling vanishes by symmetry in twisted cuprate bilayer at $\theta=45^\circ$. 


The density of states (DoS) in the chiral superconductor is still V shaped (fig \ref{fig:phase_tunnel}(b)) in the experimentally relevant energy scale. Therefore it is not possible to detect the time reversal breaking using electron spectroscopy such as STM or ARPES. Here we investigate the Kerr effect to detect the chirality. We calculate the optical conductivity tensor $\sigma_{ij}(\omega)$, signaling time-reversal breaking, plotted in fig \ref{fig:tr_45}(c,d) and from this extract the Kerr angle. The nonvanishing off-diagonal component of $\sigma_{ij}$, i.e. anomalous Hall effect, signals T breaking and is given by the anti-symmetric part of the current-current correlator $\pi_{\mu\nu}(\vec q,\omega)$,
\begin{equation}
    \sigma_h(\omega)=\frac{i}{2\omega}\lim_{\vec q\rightarrow 0} (\pi_{xy}(\vec q,\omega)-\pi_{yx}(\vec q,\omega)).
\end{equation}
The current operator is given by the procedure of minimal coupling $\vec k\rightarrow \vec k+e\vec A/\hbar$ and taking derivatives with respect to EM field $A$, $\vec j_i=\frac{1}{\Omega}\partial {H_{tbsc,\vec A}}/\partial A_i$, where $\Omega$ is the volume. Note pairing function $\Delta_k$ does not appear in the current operator since it does not involve the center of mass motion of Cooper pairs, details in appendix \ref{sigmah}.  Nonzero anomalous Hall conductivity $\sigma_h(\vec q\rightarrow 0)$ happens only if T is broken and occurs in a system that breaks Galilean invariance\cite{Kallin_2016,zhang20,denys21}\footnote{We explicitly check that for zero twist angle (single-band $d-$ wave pairing), even with nontrivial $\phi\neq 0,\pi$ and tunnel, 
$\sigma_h(\vec q\rightarrow 0)$ vanishes.}.

Due to the parton approach with the $x$ factor in interlayer tunneling, the peak in $\sigma_h$ occurs at smaller frequency $\sim 0.02eV$ than in optical measurements. The Kerr angle when the sample thickness $h\ll \lambda$ ($\lambda$ the wavelength of incident light) reads\cite{cankerr} 
\begin{equation}
    \theta_K=Re \textrm{ arctan}[\frac{\sigma_{xy}}{\sigma_{xx}+4\pi(\sigma_{xx}^2+\sigma_{xy}^2)}].
\end{equation}
At the typical frequency in fig \ref{fig:tr_45}(c), the behavior for $\sigma_{xx}$ is complicated \cite{Hwang_2007}. Though $\sigma_h$ is of same order as the Hall conductivity in \cite{cankerr}, albeit in lower frequency range, we expect the order of magnitude of $\theta_K$ is the same, i.e. $\sim \mu rad$ hence, measurable in experiments with Terahertz radiation.

 \emph{Small twist angles: Topological superconductivity and Flat band-}
At small angles, we take $t_za_0=0.16t$ to account for some homogeneous tunneling and find T is preserved. If one adds by hand an interlayer supercurrent (to model a uniform current state along the c-axis) by fixing $\phi\neq 0,\pi$, topological superconductivity emerges an in Ref. \onlinecite{pixley20} and indicated by Chern number  plotted in fig \ref{fig:phase_tunnel} (d). Chern number for $\theta=19^o,\,28^o$ reaches the maximal value of $7.7,4.5$, the non-quantization is due to numerical precision and is assigned to have $C=8, 4$ respectively. The gap for the topological superconductor is also shown fig \ref{fig:phase_tunnel} (d). 

For locating the flat band, since the $x$ factor suppresses tunneling, we expect the magic angle is suppressed from $\theta_{MA}\approx\frac{T_z}{2v_FQ}$, where $T_z$ is effective tunneling strength and $Q$ the node momentum. Indeed in fig \ref{fig:phase_tunnel}(a), the maximal possible gap $\Delta_m$ (blue curve for mixed tunneling) upon varying $\phi$ keeps increasing as twist angle $\theta$ is reduced to $9.7^o$, indicating flat band may exist for $\theta<9.7^o$, if at all. For uniform tunneling only, $\Delta_m$ reaches the maximum at $\theta\approx 13^o$. In ref.~\onlinecite{pixley20}, a magic angle around $13.8^o$ is predicted for Bi2212 from a BCS model analysis. In our parton formalism, the magic angle is reduced due to $x$ factor in the interlayer tunneling in $\mathcal H_{MF}$, as shown in the blue curve in fig \ref{fig:phase_tunnel}(a). 

For twist angle close to $45^o$, surprisingly a relatively flat portion near the top of the lowest conduction band is found with $\cos$ form tunnel. As plotted in fig \ref{fig:bandtop}, the energy scale is $\sim 5\%t$ and could result in an enhanced signal in optical measurements.

\emph { Moir\'e as a Probe of Correlation Physics-} 
We also propose to detect the pair-density-wave (PDW)\cite{agterberg2020physics} superconductor in twisted cuprate bilayer using the inter-layer pair-tunneling. We can apply a bias to give different doping in the two layers. This makes it possible to be in the regime $T>T_c^t, T<T_c^b$, where $T_c^t,T_c^b$ are the critical temperature for the uniform d wave superconductor of the two layers. Therefore the bottom layer is in the d-wave superconductor (SC) phase, while the top layer is in the pseudogap (PG) phase. Hence we have a PG-SC junction instead of the SC-SC junction.  It can be shown\cite{scalapino1970pair,lee2019proposal} that the inter-layer tunneling $I-V$ curve encodes the information of superconductor fluctuation in the top layer:

\begin{equation}
    I(V)\propto \sum_{\mathbf Q} |a(\mathbf Q)|^2 |\Delta_b|^2 \text{Im} \chi_R(\mathbf{\tilde q}-\mathbf{Q}, \omega=\frac{2e V}{\hbar})
\end{equation}
where $a(\mathbf Q)$ is the Fourier transform of the Josephson coefficient $a(\mathbf r)$. $\mathbf{\tilde q}=\frac{2 e}{\hbar}\mathbf{A_t}$, where $\mathbf{A_t}$ is the gauge field in the top layer, which can be changed by either in-plane magnetic field or a solenoid\cite{lee2019proposal}. $\chi_R(\mathbf q,\omega)$ is the retarded pair susceptibility in the top layer at momentum $\mathbf q$ and frequency $\omega$, which is a Fourier transformation of $\chi_R(\mathbf r,t)=\langle [\Delta_t(\mathbf r,t),\Delta^\dagger_t(0,0)]\rangle \theta(t) e^{-i \frac{2e}{\hbar} \int_0^{\mathbf r} \mathbf{A_t} \cdot \mathbf{dl}}$. Here $\Delta_t(\mathbf r,t)$ is the Cooper pair field in the top layer.  For untwisted bilayer, $a(\mathbf Q) \neq 0$ only for $\mathbf Q=0$. In this case the PDW fluctuation in the top layer can not couple to the uniform d-wave superconductor in the bottom layer unless we provide an unreasonably large $\mathbf{A}_t$ to compensate the momentum $\mathbf Q_{\text{PDW}}\approx (\frac{2\pi}{8a},0)$. In our twisted cuprate bilayer, we expect that the Josephson coefficient $a(\mathbf r)$ to vary within the moir\'e unit cell and we have $a(\mathbf G_M)\neq 0$, where $\mathbf G_M$ is a reciprocal vector of the moir\'e lattice.  At twist angle $\theta \approx 7.18^\circ$, we notice that $\mathbf G_M = \frac{2\pi}{a}(\sin \theta,\cos \theta-1)\approx \frac{2\pi}{a}(\frac{1}{8},-0.007) $ in the coordinate of the top layer and is close to the expected $\mathbf{Q_{\text{PDW}}}$. The small deviation may be compensated by $\tilde q=\frac{2\pi}{a}(0,-0.007)$ given by an in-plane magnetic field $B_x$. From $\frac{|\tilde q|}{2\pi/a}=\frac{e B d a}{h}$ we estimate that we need $B_x=84 $ T, assuming the inter-layer distance $d=0.95$ nm and $a=0.36$ nm. This seems to be unrealistic, but we expect to see PDW fluctuation  with a small deviation because $\text{Im} \chi_R(\mathbf q,\omega)$ should have a Gaussian form around $\mathbf{Q_{\text{PDW}}}$. Thus a small momentum deviation $\delta q=0.007 \frac{2\pi}{a}$ does not kill the signal. Alternatively we can follow Ref.~\onlinecite{lee2019proposal} to generate a larger $\mathbf{A_t}$ with a solenoid.

\emph{Conclusions-} Through a self-consistent mean field study of t-J model for twisted double-bilayer cuprates, we find a time reversal breaking but non-topological superconductor at twist angle around $45^\circ$
. Our conclusions  rely on incorporating Mott physics and form factor in the inter-layer tunneling, highlighting their importance in the modeling of twisted cuprate bilayer. With microscopic calculations, we also provide twist angle dependence of the critical Josephson current, in qualitative agreement with the recent experiment\cite{zhao21}. Finite temperature extensions of this theory and the interplay of correlation phenomena with the moire\' structure  will be interesting to explore in the future.

\emph{Acknowledgements-} We thank Philip Kim, Xiaomeng Cui and Mohit Randeria for discussions. This research was funded by a Simons Investigator award (AV), the Simons Collaboration on Ultra-Quantum Matter, which is a grant from the Simons Foundation (651440, AV), and a 2019 grant from the Harvard Quantum Initiative Seed Funding program. 

\emph{Note added-} During the finalization of the manuscript, we are aware of two preprints \cite{pixley21,franz21} which also study the twisted cuprate bilayer and discusses the critical Josephson current.

\appendix
\section{Interlayer Tunneling and Symmetries}
\label{symmetry}

\subsection{Interlayer Tunneling and Reflection}
\label{sec:formfactor}
\begin{figure}
 \captionsetup{justification=raggedright}
    \centering
        \adjustbox{trim={.35\width} {.3\height} {.05\width} {.2\height},clip}
    {\includegraphics[width=.75\textwidth]{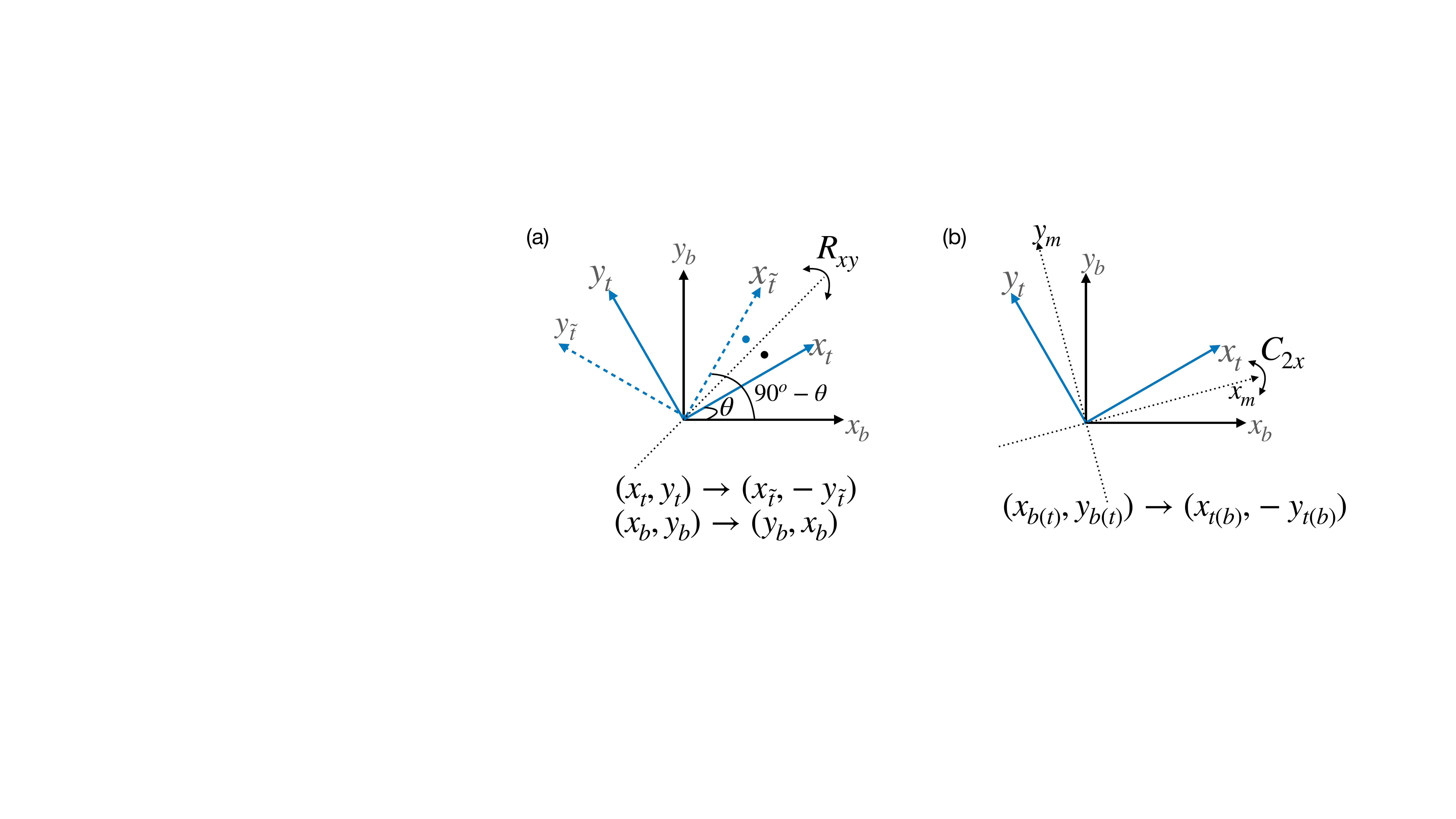}}
    \caption{(a)Illustration of $R_{xy}$ that sends $\theta\rightarrow 90^o-\theta$. The straightforward way is to write coordinates in the local reference frame of each layer, i.e.$x_{t,b},y_{t,b}$, etc. (b) $C_{2x}$ that exchanges $t,b$ layers. }
    \label{fig:rxy}
\end{figure}

We note a reflection transform $R_{xy}$ that relates twist angel $\theta\rightarrow \pi/2-\theta$ and imposes certain conditions on such interlayer tunneling form. In particular for twist angle $\theta=45^o$, $R_{xy}$ leaves the system invariant, i.e. becomes a symmetry. The reflection with respect to $x=y$ diagonal axis of the bottom layer reads
\begin{align}
    (c_{b\uparrow,i_b},c_{b\downarrow,i_b})^T\rightarrow -S_{xy} (c_{b\uparrow,Rxy(i_b)},c_{b\downarrow,Rxy(i_b)})^T,\nonumber\\
    (c_{t\uparrow,i_t},c_{t\downarrow,i_t})^T\rightarrow S_{xy} (c_{t\uparrow,Rx(i_{\tilde t})},c_{t\downarrow,Rx( i_{\tilde t})})^T,\nonumber\\
    i_t=(i_{t,x},i_{t,y}), i_{\tilde t}=(i_{\tilde t,x},-i_{\tilde t,y}),
\end{align}
where $S_{xy}=i(\sigma^x+\sigma^y)/\sqrt{2}$. $R_{xy (x)}$ denotes the reflection for spatial coordinates that sends $R_{xy}: i_x\leftrightarrow i_y$, $R_x: (i_x\rightarrow i_x,i_y\rightarrow -i_y)$, respectively and the electron operators coordinates are written in the Cartesian coordinates of the layer they are in (see fig \ref{fig:rxy}), hence the subscript $i_{t,b}$. The reflection is along diagonal direction of the bottom layer and transforms the top layer to a different twist angle $\tilde\theta=\pi/2-\theta$ with respect to the bottom layer, denoted by $\tilde b$. The electron coordinates in top layer is written in the transformed reference frame $\tilde t$ (note this is \emph{not} the reflected frame, but rather a rotated frame, hence $R_x(i_{\tilde t})$), i.e. an electron $(i_{t,x},i_{t,y})$ in the original top layer is reflected to $(i_{\tilde x},-i_{\tilde t,y})$ in the reference frame $\tilde t$. Given the $d$ orbitals are odd under reflection $R_{xy}$ for the bottom layers, while $d$ orbital wavefunction transforms to still $d$ orbitals for the top layers with the rotated reference frame, a minus sign is added for $c_b$ transforms in the first line. This reflection relates the twist angle $\theta\rightarrow \pi/2-\theta$ and the interlayer tunneling $F_{tb} (\vec r_{ij})\rightarrow -\tilde F_{tb}(R_{xy}(\vec r_{ij}))$.

In particular, for twist angle $\theta=45^o$, the system stays invariant under $R_{xy}$, imposing the constraint
\begin{equation}
\label{rxy45}
    F_{tb,45^o} (\vec r_{ij})=-F_{tb,45^o}(R_{xy}(\vec r_{ij})).
\end{equation}

We next derive the interlayer tunneling on microscopic grounds and use $R_{xy}$ to restrict the parameters. There are $2$ major processes that cause the $d$ orbital electrons to hop between top $(t2)$ and bottom $(b1)$ layers: first one is mediated by $s$ orbitals and second one is a direct tunneling from $d$ orbitals active in the $t-J$ model. The first one consists further of $3$ steps: $d$ electrons hop to $s$ orbitals in the same CuO plane, i.e. within the $t2,b1$ layers, with a form factor $\cos k_x-\cos k_y$ and $s$ orbitals hop between $t2,b1$ layers, with exponential decay $g_s(i,j)=e^{-(l_{ij}-l_d)/\rho_s}$ (the $s$ orbital distance $l_{ij}$ and $l_d$ the distance between top $t2$ and bottom $b1$ layers). The second one assumes a form of
\begin{equation}
    \label{uniformtn}
F_{tb,2}(\vec r_{ij})=a_0 e^{-(l_{ij}-l_d)/\rho_d} c^\dagger _{i,t2} c_{j,b1}.
\end{equation}
 $\rho_{s,d}$ is the spatial extent for $s,d$ orbitals, respectively.

At $\theta=45^o$, the first process obeys the reflection $R_{xy}$ automatically, since the form factor $\cos k_x-\cos k_y$ takes into account the $d$ orbital symmetries. The second process is even for $l_{ij}\rightarrow R_{xy}(l_{ij})$ while $R_{xy}$ requires it being odd in eq \eqref{rxy45}. Hence we conclude the $a_0=0$ for $\theta=45^o$. This is intuitive from the $d$ orbital symmetries: at $\theta=45^o$, $d$ orbitals in bottom layer is odd under $R_{xy}$, while $d$ orbitals in top layer is even under $R_{xy}$, so they cannot mix directly.

For the first process, the tunneling form reads in real space as
\begin{align}
\label{ftbform}
    F_{tb,1}(\vec r_{ij})=\frac{1}{2}\sum_{\hat h=x,y,\hat h'=x',y'} \xi_{\hat h}\xi_{\hat h'} g_s(r_{i\pm\hat h,j\pm\hat h'}),\nonumber\\    \xi_{\hat x}=1,\xi_{\hat y}=-1,g_s(i_{t2},j_{b1})=e^{-(l_{ij}-l_d)/\rho_s},
\end{align}
where we write the $\cos kx-\cos ky$ form as neighboring hopping with direction-dependent phase $\xi_{\hat h}$. So $d$ electrons hop to $s$ orbitals in neighboring ions, and $s$ orbitals tunneling between layers and this effectively realizes tunneling of $d$ electrons between $t2,b1$ layers.

\subsection{Relation between $(\theta,\phi)$ and $(\pi/2-\theta,\pi-\phi)$}
\begin{figure}
 \captionsetup{justification=raggedright}
    \centering
        \adjustbox{trim={.0\width} {.1\height} {.0\width} {.1\height},clip}
    {\includegraphics[width=.4\textwidth]{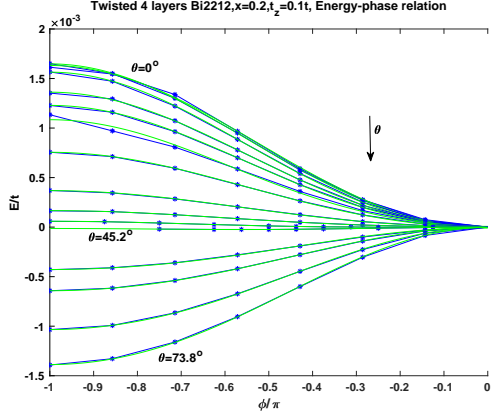}}
    \caption{The energy phase relations for various twisting angles $\theta$ (increasing from top to bottom), at $t_z=0.1t,a_0=0$. At $\theta=43.6^o,45.2^o$, T is broken, otherwise $\phi=0 (\pi)$ for $\theta<(>)45^o$, respectively.}
    \label{fig:ene4f}
    \end{figure}

We note the transformation $R_{xy}$ that sends the system with twist angle $\theta$ to $\pi/2-\theta$.
For the parton Hamiltonian $\mathcal H_{MF}$, this reflection sends
\begin{align}
    \Delta_{i,i+\hat x(\hat y),b}\rightarrow \Delta_{i,i+\hat y(\hat x),b}\nonumber\\
    \Delta_{i,i+\hat x(\hat y),t}\rightarrow \Delta_{i,i+\tilde x(\tilde y),t},
\end{align}
and leaves homogeneous mean-field $\chi$'s invariant. The phase difference $\phi$ therefore becomes $\phi-\pi$ given the d-wave pairing ansatz. For the special case of $\theta=45^o$, the reflection leaves the two layers unchanged, with pairing phase difference $\phi\rightarrow \phi-\pi$, further related to $\pi-\phi$ by T. Hence the reflection symmetry $R_{xy}$ means the mean-field energy vs $\phi$ should be symmetric for $\phi,\pi-\phi$, which our results for $\theta=45.2^o$ approximately obeys in fig \ref{fig:ene4f}. 

\subsection{Other symmetries}
The system has an anti-unitary chiral symmetry $\mathcal C$:
\begin{eqnarray}
\mathcal C: c_{is,p}\rightarrow \mathcal K \sigma^z_{ss} c^\dagger_{is,p}\nonumber\\
\mathcal C^\dagger \mathcal H_{tbsc} \mathcal C=-\mathcal H_{tbsc},
\end{eqnarray}
where the spin-dependent factor $ \sigma^z_{ss} $ takes care of the singlet pairing terms to become negative upon the transformation.

We focus on commensurate twist, i.e. when the enlarged, exact, square unit cell contain a finite number of sites, characterized by a twist vector $(n,m)$ with the angle $\theta=2 arctan(n/m)$ as the twist angle. In this case, a four-fold rotation $C_4$ and two-fold rotation while interchanging $t,b$ layers $C_{2x}$ are present, acting as
\begin{eqnarray}
C_4: c_{is,p}\rightarrow \sigma^z_{ss} c_{C_4(i)s,p}\nonumber\\
C_{2x}:     (c_{b\uparrow,i_b},c_{b\downarrow,i_b})^T\rightarrow - (c_{t\uparrow,Rx(i_b)},c_{t\downarrow,Rx(i_b)})^T,\nonumber\\
    (c_{t\uparrow,i_t},c_{t\downarrow,i_t})^T\rightarrow - (c_{b\uparrow,Rx(i_{t})},c_{b\downarrow,Rx( i_{t})})^T,\nonumber\\
    R_x((i_x,i_y))=(i_x,-i_y),
\end{eqnarray}
where $C_4(i)$ is the $C_4$ rotated site of the original site $i$. $C_{2x}$ (illustrated in fig \ref{fig:rxy}(b))looks like a reflection when restricting to the $2D$ cuprate planes, where $R_x(i)=(i_x,-i_y)$ is the transformed coordinates of $C_{2x}$, written in the \emph{transformed} layer $\epsilon_{t,b}=b,t$ reference frame.

\section{Some Band Structure Details}
\label{bandstr}
Taken from Ref \cite{Bansil}, in Bi2212 there are two kinds of hoppings, within each bilayer $t_{bi} = 0.3t$ and between bilayers $t_z=0.1 t$. The `z' dispersion is stated to be:
\begin{eqnarray}
    E_z &=& T_z(k_\parallel,\,\cos k_z c/2) \left \{\frac{1}{4} [\cos(k_x a) - \cos  (k_ya)]^2+\tilde a_0\right \}  \nonumber \\
    T_z &=& \pm \sqrt{t_{bi}^2 + A^2 +2 t_{bi} A \cos k_z c/2}\nonumber \\
    A &=& 4\tilde t_z \cos k_x a/2 \cos k_y a/2 
\end{eqnarray}
and $\tilde a_0 = 0.4$ which gives some tunneling along the nodal direction. In the main text, we have $a_0=4\tilde a_0$ to account for the factor $4$ in expression for $A$ and hence $a_0=1.6$.

Note there is some statement that the tunneling $t_z$ which connects different bilayers has some structure corresponding to displacements (corresponding to A above). We take the geometry as no displacement between $2$ layers and find a displacement of site close to $(a/2,a/2)$ for $\theta\sim 45^o$. 

For the twisted setting, we modify the factor $\cos(k_x a) - \cos  (k_ya)]^2$ in $E_z$ to $(c_{x_t}-c_{y_t}) (c_{x_b}-c_{y_b})$ where $c_{x(y)}\def \cos(k_{x(y)}a)$ and we write the two factors in reference frame of top, bottom layers, hence the subscript $t,b$. This gives rise to the form factor in eq \eqref{ftbform} in real space. The uniform $a_0$ factor corresponds to the second process $F_{tb,2}$ mentioned  in eq \eqref{uniformtn}.

\section{Details on numerical iteration}
\label{app:num}
We iterate for a self-consistent solution for $\chi,\Delta,x$'s, i.e. substitute the order parameters by the expectation value of the corresponding operators,
 \begin{eqnarray}
 \chi_{ij}=\sum_{k,n} n_f(\epsilon_n) \langle \psi_{k,n}| f_{is}^\dagger f_{is}|\psi_{k,n}\rangle,\nonumber\\
 \Delta_{ij}=\sum_{k,n}\sum_{ss'} n_f(\epsilon_n) \epsilon_{ss'}\langle \psi_{k,n}|f_{is}f_{js'}|\psi_{k,n}\rangle,\nonumber\\
 x_i=\sum_{k,n} n_f(\epsilon_n) \langle \psi_{k,n}| (1-\sum_sf_{is}^\dagger f_{is})|\psi_{k,n}\rangle
 \end{eqnarray}
 where $n$ labels the eigenstate at a certain momentum $\mathbf k$, and $n_f(\epsilon_n)$ is the Fermi function for the eigenenergy $\epsilon_n$.
 
 In numerics, we switch to an equivalent model setting where the pairing is enforced to be real numbers, and phase difference $\phi$ enters through a phase for the interlayer tunneling $F_{tb}\rightarrow F_{tb}e^{i\frac{\phi}{2}}$. By a phase rotation of the spinon in bottom layer $f_{is,b}\rightarrow f_{is,b}e^{-\frac{\phi}{2}}$, one comes back to real $F_{tb}$'s with pairing difference $\phi$. This transformation allows us to obtain self-consistent solutions with a fixed $\phi$, likely with a nonvanishing interlayer supercurrent.  We also find that if one takes a simplified model of twisted bilayer cuprates,with only $t,b$ layers, the results stay qualitatively invariant compared with twisted double-bilayer ($4$ layers in total) settings, i.e. regarding time-reversal breaking. The gap obtained in $2$ layers is larger (roughly by a factor of $2$) than that for twisted double-bilayers.  The Chern number results are different for $2,4$ layers calculation in some cases, e.g. at small twist angles with nontrivial $\phi$, $2,4$ layers give $C=4,8$, respectively. The results of hall conductivity are obtained in a twisted bilayer (2 layers) setting.
 
 We focus on commensurate twist, i.e. when the enlarged, exact, square unit cell contain a finite number of sites, characterized by a twist vector $(n,m)$ with the angle $\theta=2 arctan(n/m)$ as the twisted angle.

   \section{Continuum model and gap}
  \label{continuum}
  We use a linearized model near the Dirac cone of $d-$wave SC and add interlayer tunnel\cite{pixley20} to see how interlayer tunneling affects the size of gap. To further simplify, we consider $2-$ band models in the top and bottom cuprates, respectively. The Dirac Hamiltonian for one of the Dirac fermions, for the $2$ layers read ($\sigma$ acts in Nambu spinor space, $\tau$ acts in layer space) in the basis $(\psi_{k\uparrow,t},\psi_{-k\downarrow,t}^\dagger,\psi_{k\uparrow,b},\psi_{-k\downarrow,b}^\dagger)^T$,
  \begin{eqnarray}
  H_{Dirac}(\vec k)=(v_F k_{\perp}+\mu)\sigma^3+v_{\Delta} k_{\parallel}\sigma^1+v_{\Delta}|Q| \sigma^1\tau^3,
  \end{eqnarray}
  where $k_{\parallel,\perp}$ represents momenta parallel, perpendicular to the Fermi surface, $v_{F,\Delta}$ are velocities for Dirac fermions in the normal state, from gap function, respectively. $2\vec Q$ is the twist vector, $2\vec Q\approx\theta\hat z\times \vec K$, where $\vec K(=(\pm\pi/2,\pm \pi/2))$ is the momentum of Dirac point. The last term accounts for the mismatch of the Dirac points momenta in two layers due to the twist, i.e. we are expanding around $\vec K+\vec Q$.
  \begin{widetext}
  The tunneling with a phase $\phi/2$, equivalent to a pairing phase difference $\phi$, reads in the low-energy space,
  \begin{eqnarray}
  H_{tunnel}(\vec k)=T_z(\vec k)(\cos \frac{\phi}{2} \sigma^3\tau^1+\sin \frac{\phi}{2} \tau^2),
  \end{eqnarray}
  where $T_z(\vec k)=t_zxF_{tb}(\vec k)$ is the effective tunneling matrix. The eigen energy is
  \begin{eqnarray}
      E^2&=&(v_Fk_\perp+\mu)^2+(v_\Delta k_\parallel)^2+(v_\Delta Q)^2+(T_z(k))^2\nonumber\\
      \pm &2&\sqrt{(v_Fk_\perp+\mu)^2(T_z(k))^2+(v_\Delta Q)^2((T_z(k)\cos\phi/2)^2+(v_\Delta k_\parallel)^2)+(v_\Delta k_{\parallel}T_z(k)\sin\phi/2)^2}.\nonumber
  \end{eqnarray}
 The minimum of $E$ requires $d E^2/dk_\perp=0,d E^2/dk_\parallel=0$. The first derivative is satisfied at $v_Fk_\perp+\mu=0$. 
We have minimum of $|E|$ at the minimum of the following expression, 
  \begin{eqnarray}
(v_\Delta k_\parallel)^2+(v_\Delta Q)^2+(T_z(k))^2-2\sqrt{(v_\Delta Q)^2((T_z(k)\cos\phi/2)^2+(v_\Delta k_\parallel)^2)+(v_\Delta k_{\parallel}T_z(k)\sin\phi/2)^2+(v_\Delta^2Qk_\parallel)^2},\nonumber
  \end{eqnarray}
  which is always positive for $\phi\neq 0,\pi$. The minimum is reached when 
  \begin{eqnarray}
  (v_\Delta k_{\parallel,m})^2((v_\Delta Q)^2+(T_z(k)\sin \phi)^2)=((v_\Delta Q)^2+(T_z(k)\sin \phi)^2)^2-(T_z(k)\cos \phi)^2(v_\Delta Q)^2,
  \end{eqnarray}
 leading to $k_{\parallel,m}$ close to $Q$ given $v_\Delta Q\gg T_z(k_{\min})$, with the minimal energy
  \begin{equation}
      E_{min}\approx \frac{(T_z(k_{min}))^2\sin\phi}{2\sqrt{(v_\Delta Q)^2+[(T_z(k)\sin \phi/2]^2}}.
  \end{equation}
  Crucially for not so large $\mu,T_z$, $\vec k_{min}$ lies close to the original Dirac point $\vec K$ where $F_{tb,1}$ vanishes. Hence the gap should be vanishingly small if the tunneling is purely $F_{tb,1}(k)=(\cos k_{x,t}-\cos k_{y,t})(\cos k_{x,b}-\cos k_{y,b})$.
  
  Another saddle point solution with nonzero $v_Fk_\perp+\mu$ gives a saddle point energy $E=2|v_\Delta Q\sin \frac{\phi}{2}|$, and at not so small twist angle, is larger than the above $E_{min}$ given $T_z$ is suppressed by a factor of $x\sim 0.1$.
  
  From this crude analysis we see how the $\cos$ form factor leads to vanishing gap. In reality, higher-order processes  that scatter between momenta $\vec k,\vec k+\vec K_{moire}$ will also contribute.
  \end{widetext}

\section{Comparison with BCS calculation results}
\label{comparebcs}
The $x$ factor from parton approach and $\cos$ form factor in tunneling in t-J calculation result in several key differences from the BCS calculation \cite{franz}:

    \begin{figure}
 \captionsetup{justification=raggedright}
    \centering
        \adjustbox{trim={.1\width} {.0\height} {.1\width} {.1\height},clip}
    {\includegraphics[width=.5\textwidth]{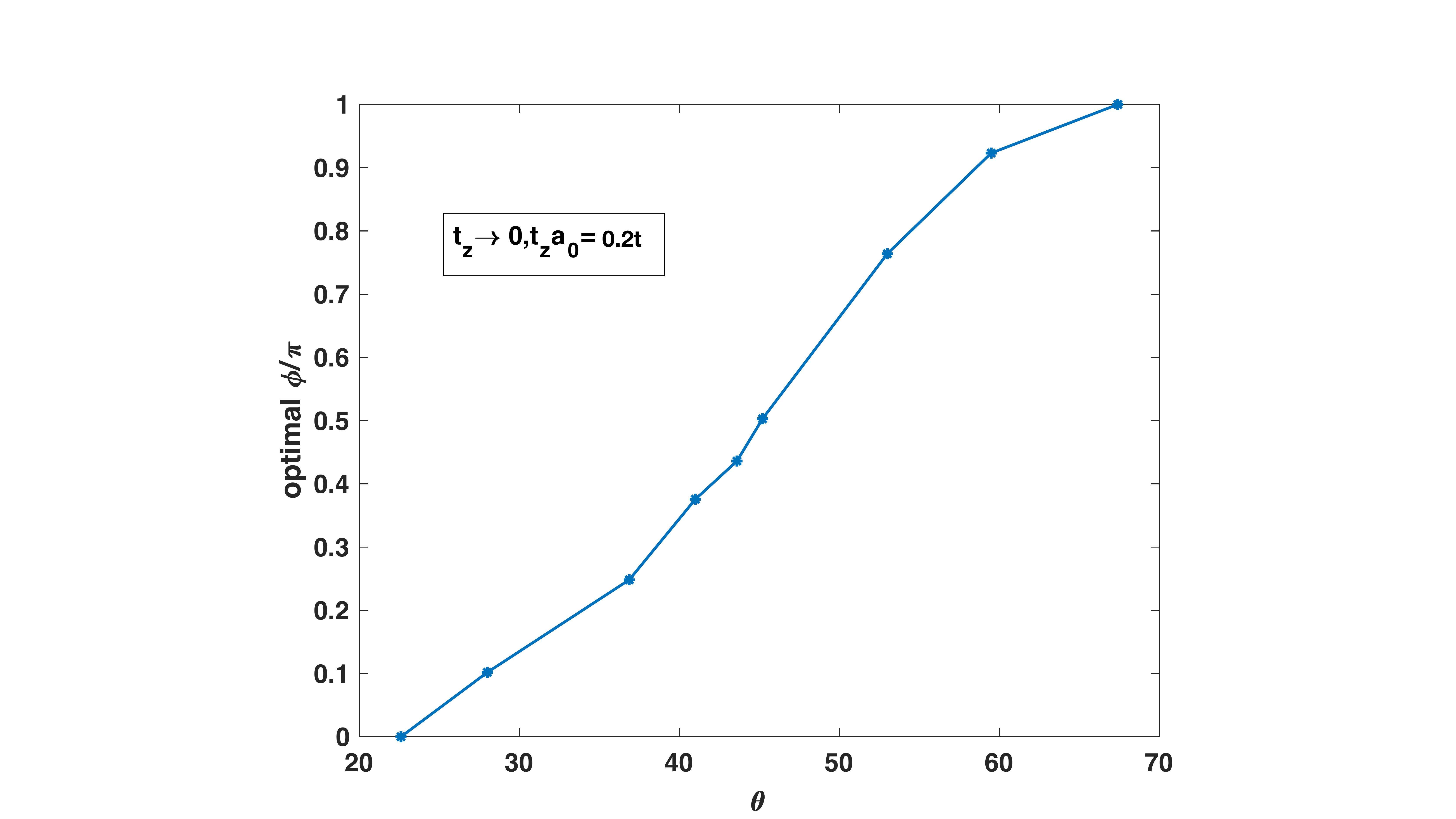}}
    \caption{The optimal phase difference $\phi$ as a function of twisting angle with only uniform tunneling $t_z\rightarrow 0,t_za_0=0.2t$.The results agree with those in \cite{franz} qualitatively.}\label{fig:phiuniform}
    \end{figure}
    
1. The $d-$ wave solution with $\phi\neq 0$ survives to $t_z$ as large as $0.3t$(fig \ref{fig:tr_45}(d)), while in BCS calculation $d+is$ or other mean-field solution set in at $t_z\approx0.15t$. 

2.The $\theta$ with spontaneous T breaking falls in a narrower window than in BCS calculation,i.e. $43^o<\theta<47^o$,since here a commensurate twist  $\theta=41^o$ already yields $\phi=0$. 

3.The minimal gap resulting from interlayer tunneling orders of magnitudes smaller $\Delta\approx0.0004t$ for $\theta\approx45^o$, while BCS calculation gives a gap $\Delta\approx0.02t$.

4. Despite the T breaking, the chern number is $0$ identically.For small twist angle $\theta<40^o$, a uniform tunneling component is allowed by symmetry and we take $t_za_0=0.16t$ from \cite{Bansil}. We found that T is preserved. If we add by hand a nonzero pairing phase difference $\phi$, the system enters topological superconductivity phase witnessed by chern number $C=8,4$ plotted in fig \ref{fig:tr_45}(d),as predicted by \cite{pixley20}. 

\begin{figure}
 \captionsetup{justification=raggedright}
    \centering
        \adjustbox{trim={.1\width} {.5\height} {.5\width} {.1\height},clip}
    {\includegraphics[width=.75\textwidth]{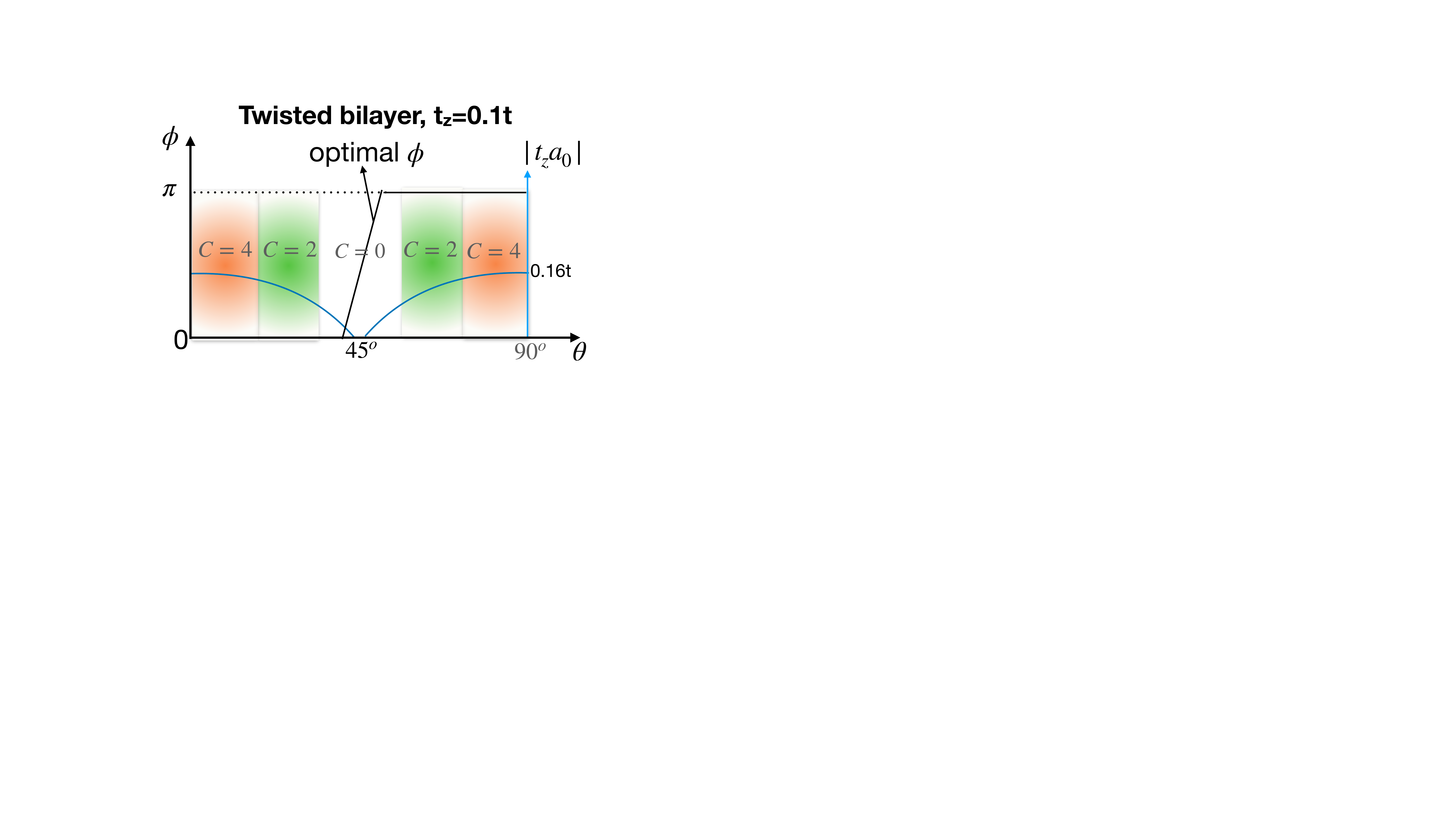}}
    \caption{A schematic phase diagram for twisted bilayer Bi2212 calculation. The results agree  qualitatively with those of twisted double-bilayer Bi2212, except Chern numbers are halved.}
    \label{fig:phase2}
\end{figure}
    
We have also calculated cases with only an exponentially decaying and homogeneous tunneling t-J calculation,i.e. setting the  process with form factor $\cos kx-\cos ky$, $F_{tb,1}=0$ identically. Operationally it is achieved in $\mathcal H_{MF}$ by taking $t_z\rightarrow 0,t_za_0=0.2t$. Then we recovered results in ref\cite{franz}: Gap $\Delta\approx 0.01 t$ for $t_za_0=0.2t,\theta=43.6^o$(gap as $\theta$ varies is plotted in fig \ref{fig:phase_tunnel}(a)), with a much larger $J_c=7000 kA/cm^2$ ($J_c=800kA/cm^2$ for the same parameters except setting $a_0=0,t_z=0.1t$ instead). 
The T breaking twist angle range is significantly larger as shown in fig\ref{fig:phiuniform}. This resembles \cite{franz} results with exponential decaying tunneling for $\Delta=0.022t,\phi=0.5\pi$ at $\theta=43.6^o$.

\section{Details on Hall conductivity calculations}
\label{sigmah}
Expand in the eigenbasis of the mean-field Hamiltonian $\mathcal H_{MF}$, the hall conductivity reads ($\eta\rightarrow 0$)
\begin{align}
    \label{sh}
    \sigma_h(\omega)=\frac{e^2}{i\hbar^2\omega \Omega}\sum_{n,m}(F(E_m)-F(E_n))\frac{\langle n|\frac{\partial H}{\partial k_x}|m\rangle\langle m|\frac{\partial H}{\partial k_y}|n\rangle }{\hbar\omega+i\eta+E_m-E_n}\nonumber\\-[k_x \leftrightarrow k_y],
\end{align}
where $F(E)$ is the Fermi function.

A simplification gives,
\begin{align}
\label{etaapr}
    \sigma_h(\omega)=\frac{e^2}{i\hbar \Omega}\sum_{E_n>E_m}(F(E_m)-F(E_n))\nonumber\\ \frac{\textrm{Im}\langle n|\frac{\partial H}{\partial k_x}|m\rangle\langle m|\frac{\partial H}{\partial k_y}|n\rangle }{(\hbar\omega+i\eta)^2-(E_m-E_n)^2},
\end{align}
in practice to regulate the singularity when $E_n-E_m=\omega$, we take the small parameter $\eta=0.004t$ as a damping factor. 

\begin{figure}
 \captionsetup{justification=raggedright}
    \centering
            \adjustbox{trim={.1\width} {.1\height} {.1\width} {.05\height},clip}
    {\includegraphics[width=.4\textwidth]{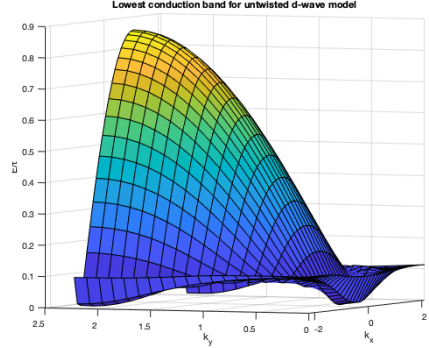}}
    \caption{The band dispersion for untwisted d-wave models with maximal gap $\approx0.2t$ along the Fermi surface for reference.}
    \label{fig:band11}
\end{figure}

\begin{figure}
 \captionsetup{justification=raggedright}
    \centering
        \adjustbox{trim={.1\width} {.2\height} {.1\width} {.1\height},clip}
    {\includegraphics[width=.5\textwidth]{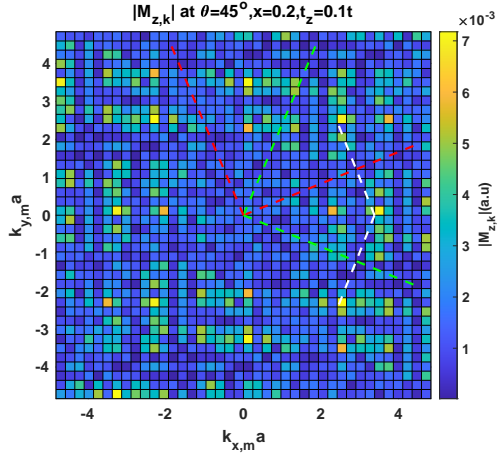}}
    \caption{Magnetization density $M_z$ (along c axis) from interlayer currents at $\theta=45.2^o$. Dashed lines are guide for the eyes of the two cuprate layers Brillouin zones. The magnetization respects $C_{2x}$ and (approximate) $R_{xy}$. The scattering between the bright yellow spots located around e.g. $(2.4,2.4),(3.1,0)$ etc will give peaks along reciprocal vector direction of the two cuprate layers,i.e. parallel to one of the dashed lines. This is illustrated by the white lines connection the yellow spots. In-plane Magnetization is an order smaller than $|M_z|$. }
    \label{fig:mkz}
\end{figure}

\begin{figure*}
 \captionsetup{justification=raggedright}
    \centering
        \adjustbox{trim={.03\width} {.12\height} {.08\width} {.22\height},clip}
    {\includegraphics[width=.8\textwidth]{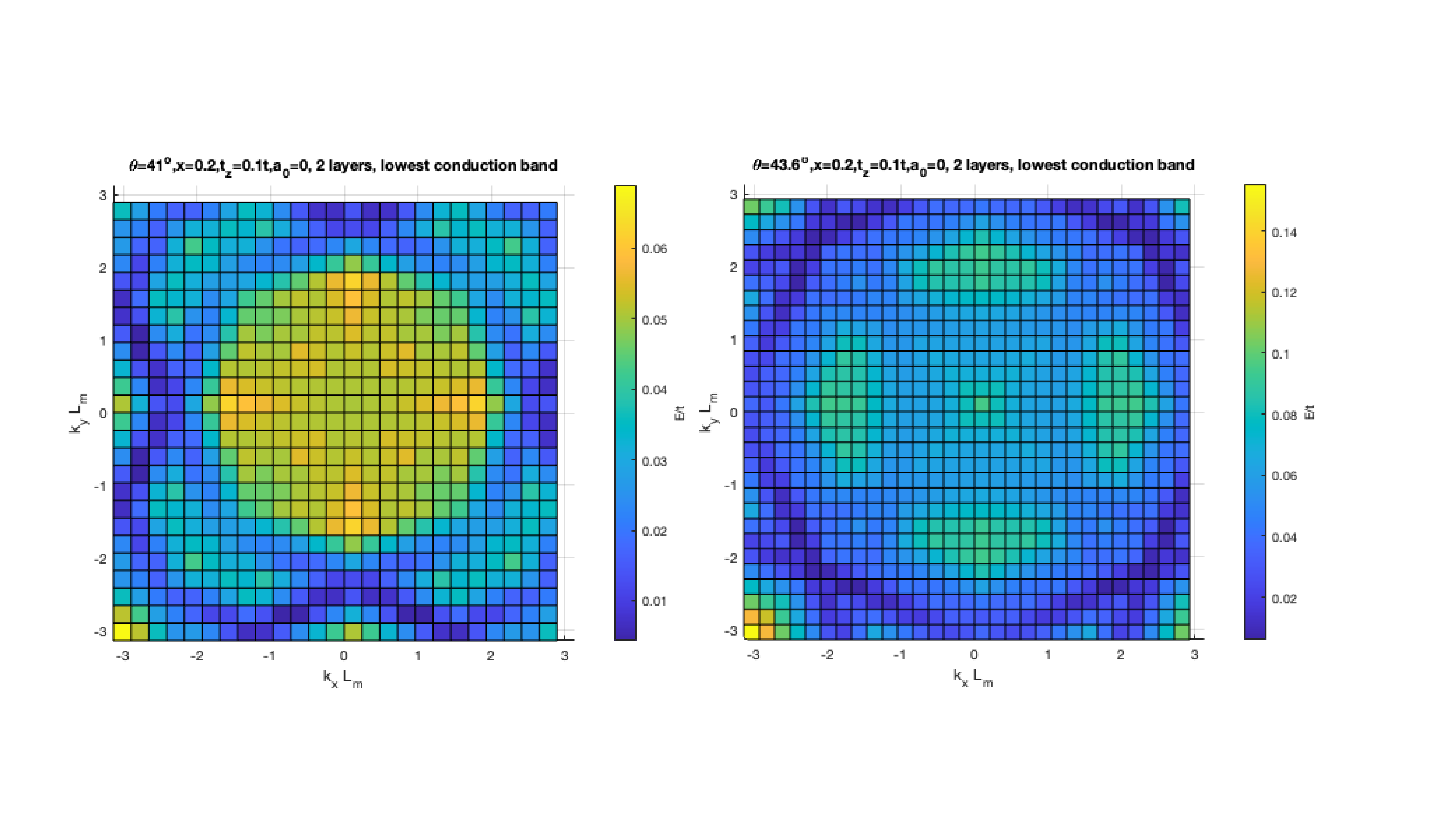}}
    \caption{Lowest conduction band at $\theta=41^o,43.6^o$ for the left and right panel, respectively, with $t_z=0.1t,a_0=0$. Although no flat band in the lowest energy, we find a relatively flat band top at $E\approx0.05,0.06 t$, respectively, which is signaled by a peak in optical conductivity measurement around $\omega=400,500 cm^{-1}$.}
    \label{fig:bandtop}
\end{figure*}

%

\end{document}